\documentclass[preprintnumbers]{revtex4}

\usepackage{graphicx}
\def\beq{\begin{equation}}
\def\eeq#1{\label{#1}\end{equation}}
\def\eeqn{\end{equation}}
\def\beqa{\begin{eqnarray}}
\def\eeqa#1{\label{#1}\end{eqnarray}}
\def\eeqan{\end{eqnarray}}
\def\CR{\nonumber \\ }
\def\leqn#1{(\ref{#1})} 

\def\half{{1 \over 2}}
\def\sstw{\sin^2 \theta_W}        
\def\cstw{\cos^2 \theta_W} 
\def\mz{M_Z}

\newenvironment{mydesc}
               {\begin{list}{} %
                {\itemsep=0pt \topsep=0pt \partopsep=0pt \parsep=1pt %
                 \leftmargin=19pt \labelwidth=15pt %
                 }} 
               {\end{list}}
                                          
\begin{document}
\bibliographystyle{revtex}

\preprint{ LBNL-49222 }

\title{Exploring New Physics Through Contact Interactions in Lepton Pair Production 
at a Linear Collider}



\author{Gabriella P\'asztor} 
\email[]{Gabriella.Pasztor@cern.ch}
\altaffiliation{On leave of absence from KFKI RMKI, Budapest, Hungary.}
\affiliation{CERN, CH-1211 Geneva 23, Switzerland}

\author{Maxim Perelstein}
\email[]{meperelstein@lbl.gov}
\affiliation{Theory Group, Lawrence Berkeley National Lab, Berkeley, 
CA 94720, USA}


\date{\today}

\begin{abstract}
If a contact interaction type correction to a Standard Model process is
observed, studying its detailed properties can provide information on the
fundamental physics responsible for it. Assuming that such a correction has
been observed in lepton pair production at a 500 GeV $-$ 1 TeV linear 
collider, we consider a few possible models that could explain it, such as 
theories with large and TeV-scale extra dimensions and models with lepton
compositeness. We show that using the measured cross-sections and angular 
distributions, these models can be distinguished with a high degree of
confidence.    

\if
We study lepton pair production at a 500 GeV $-$ 1 TeV linear collider  
to extract information on the fundamental theory above the
TeV scale via contact interactions. 
We consider models with large and TeV-scale extra dimensions
and lepton compositeness and show that using the measured cross-sections and
angular distributions these models can be distinguished whenever a
significant deviation from the Standard Model is observed.
\fi
\end{abstract}


\maketitle

\section{Introduction}
All known solutions to the gauge hierarchy problem of the Standard Model
(SM) require the appearance of new particles at energy scales around 
1 TeV. 
It is not guaranteed, however, that these new particles can be produced   
directly at the proposed 500 GeV linear collider (LC). Only for
supersymmetric theories are there strong arguments that at least some 
superpartners should be kinematically accessible at such a 
collider~\cite{Abe:2001wn}. In the case of composite Higgs models and models 
with extra dimensions, the situation is far less certain. It is possible
that all the new states predicted in these theories are too heavy and cannot 
appear in the final state at a 500 GeV LC. In fact, for models
with large extra dimensions~\cite{ADD}, current experimental constraints most 
likely rule
out the possibility that string Regge excitations could be lighter than 500 
GeV. In this case, the only direct effect of extra dimensions would be the 
enhanced rate of events with missing energy due to graviton emission. These 
events, however, provide only very limited amount of information about the 
fundamental theory. Moreover, this signature could be mimicked by gravitino 
emission processes in certain supersymmetric models, so one would need 
additional handles to disentangle the underlying physics~\cite{GPW}. In this 
situation, it is important to look for indirect effects of new 
physics, that is, the effects of new heavy particles appearing as virtual
states. For example, processes such as Bhabha scattering or other lepton pair
production, 
\beq
\mathrm{ e^+ e^- \rightarrow e^+ e^-}, \mu^+ \mu^-, \tau^+\tau^- 
\eeq{eq:bhabha}
could receive an additional contribution from the exchange of a heavy
state $X$. Because such additional contributions come from short-distance 
physics and do not possess poles in the accessible range of any
kinematic variables, they 
are referred to as {\it contact interactions.}
By carefully examining the total cross section and angular 
distribution of these processes, 
it should be possible to not only find deviations 
from the Standard Model, but also gain some information about the nature of 
the state $X$, such as its spin and couplings. 

In this report, we will assume that the cross section of process 
\leqn{eq:bhabha} was found to deviate from the Standard Model prediction. We 
will then consider several possible explanations for this deviation, such as 
models with lepton substructure, models with TeV-scale strings, and models in 
which gauge fields can propagate in the extra dimensions. Our main goal is to 
determine 
how well one can discriminate between these possibilities, given the 
measurement of the total cross section and angular distributions of the 
final-state particles. 

\section{Models with Contact Interactions}
The unpolarised
cross section formula for Bhabha scattering can be written in the form
\beq
{d \sigma\over d \cos\theta} = {\pi\alpha^2\over 2 s} \left[
  u^2  (|A_{LL}|^2 + |A_{RR}|^2 ) + 2 t^2 |A_{RL,s}|^2  + 2s^2 |A_{RL,t}|^2
        \right]\ ,
\eeq{eq:BH1}
where
\beqa
   A_{LL} &=& {1\over s} + {1\over t} + {(\half-\sstw)^2 \over \sstw\cstw}
                 \left( {1\over s-\mz^2} + {1\over t-\mz^2}\right)
                      +  \Delta_{LL} \ , \CR
   A_{RR} &=& {1\over s} + {1\over t} + {\sstw \over \cstw}
                 \left( {1\over s-\mz^2} + {1\over t-\mz^2}\right)
                      +  \Delta_{RR}\ , \CR
   A_{RL,s} &=& {1\over s}  -  {(\half-\sstw) \over \cstw}
                 {1\over s-\mz^2}
                      + \Delta_{RL,s} \ , \CR
   A_{RL,t} &=& {1\over t}  -  {(\half-\sstw) \over \cstw}
                 {1\over t-\mz^2}
                      + \Delta_{RL,t}  \ ,
\eeqa{eq:BH2}
and the $\Delta_a$ functions represent the contact interaction corrections 
coming from TeV-scale physics. For this study we have considered the 
following models:
\begin{mydesc}
\item[Models with composite leptons]~\cite{Eichten:1983hw}, where the 
contact interaction terms are given by
\beqa
   \Delta_{LL} = 2{\eta_{LL}\over \alpha \Lambda^2}, \hskip2cm
   \Delta_{RR} =  2 {\eta_{RR}\over \alpha \Lambda^2}, \hskip2cm
  \Delta_{RL,s}= \Delta_{RL,t} =  {\eta_{RL}\over \alpha \Lambda^2}.
\eeqa{eq:BH3}   
Here $\eta_a = \{+1, 0, -1\}$ parametrise the helicity structure of the 
contact interactions, and $\Lambda$ is the scale of compositeness. We 
will study two possibilities: 
\begin{mydesc}
\item[(VV)] the vector-vector model with $\eta_{LL}=\eta_{RR}=\eta_{RL}=+1$,
\item[(AA)] the axial-axial model with $\eta_{LL}=\eta_{RR}=-\eta_{RL}=+1$. 
\end{mydesc}
\item[Models with large extra dimensions]
have two sources from which contact 
interactions may arise. The first contribution is from the 
virtual effect of string Regge excitations of the photon and the $Z$ boson. 
This has been computed in~\cite{Cullen:2000ef} using a simple string toy 
model. The corrected Bhabha scattering cross section is given by
\beqa
{d \sigma\over d \cos\theta}(e^+e^-\rightarrow e^+e^-) =
 {d \sigma\over d \cos\theta}\biggr|_{SM}  \,\times\,\left( 1 - {\pi^2 \over 
3} {st \over M_s^4} + \ldots \right),
\eeqa{eq:string} 
where $M_s$ is the string scale.
The second contribution comes from virtual graviton exchange, 
and was analysed in~\cite{Giudice:1999sn,Hewett:1999sn,Rizzo:1999sn}. 
This effect could be sizable 
because of the large number of Kaluza-Klein (KK) modes of the graviton that
contribute. The $\Delta_a$ functions in this case are given by
\beqa
   \Delta_{LL}= \Delta_{RR} = {\lambda\over \pi\alpha M_H^4} \left[
                 (u + {3\over 4} s) + (u + {3\over 4} t)\right],\hskip1.7cm\CR
   \Delta_{RL,s} =  - {\lambda\over \pi\alpha M_H^4}
                         (t + {3\over 4} s),\hskip2cm
  \Delta_{RL,t} =  - {\lambda\over \pi\alpha M_H^4}
                         (s + {3\over 4} t) \ ,        
\eeqa{KKgrav}
where $M_H$ is the quantum gravity scale as defined in~\cite{Hewett:1999sn}.
Here we will study two models: 
\begin{mydesc}
\item[(SR)] The String Regge model, where the  
contribution of the Regge states is dominant, as is necessarily the case if 
physics at the TeV scale is described by weakly coupled string theory. 
\item[(KK+, KK$-$)]
The KK graviton model with $\lambda=+1$ or $-1$, where we assume
that the Regge contribution is for some reason suppressed, and the virtual
graviton exchange dominates. 
\end{mydesc}
\item[Models with TeV-scale extra dimensions (TeV)] may allow
the Standard Model gauge bosons to propagate in the additional dimensions.
In this case, contact interactions arise from the exchange
of virtual KK excitations of the photon and the $Z$ boson. 
Using the formalism of \cite{Rizzo:2000br}, we obtain
\beqa
   \Delta_{LL} = -\half\,{1 \over \cstw \sstw}\,{V\over M_W^2}, \hskip2cm
   \Delta_{RR} =  -2\,{1 \over \cstw}\,{V\over M_W^2},\CR
  \Delta_{RL,s}= \Delta_{RL,t} = 
                 -\half\,{1 \over \cstw}\,{V\over M_W^2}, \hskip3cm
\eeqa{eq:KKgauge}
where in the case of one extra dimension $V$ is directly related to the 
compactification scale $M_c$:
\beq
V = {\pi^2 \over 3}\,{M_W^2 \over M_c^2}.
\eeq{Vfor1}
For more than one extra dimension, the relation between $V$ and $M_c$ depends
on the details of the TeV-scale physics, and it is more useful to work in terms
of $V$ itself. In this study we give all the results in terms of $M_c$. 
\end{mydesc}

Analogous formulas can be obtained for $\mu^+\mu^-$ and $\tau^+\tau^-$ 
final states. Since phenomenology of string models with multiple generations
has not been studied in detail, we will not consider the effects of Regge
states in these channels. 

\section{Analysis}

\begin{figure}[htbp]
\centerline{ 
\includegraphics[width=14cm]{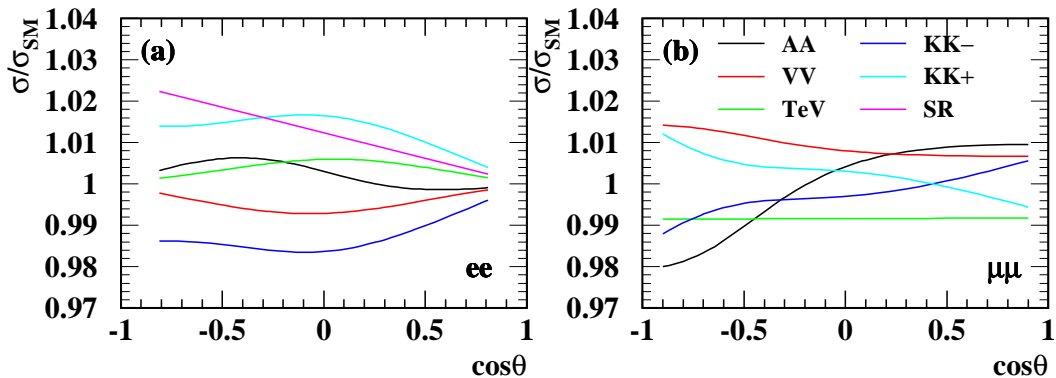}
}
\caption{
Ratio of the predicted new physics cross-section to the SM cross-section
for (a) electron and (b) muon
pair production as a function of the lepton polar
angle for the models
AA ($\Lambda$=71 TeV), VV ($\Lambda$=85 TeV), TeV ($M_c$=14 TeV), 
KK+ ($M_H$=3.4 TeV), KK$-$ ($M_H$=3.4 TeV)
and SR ($M_s$=1.7 TeV) at a 500 GeV LC. 
The scale parameters have been
chosen to be at the sensitivity reach with
100 fb$^{-1}$ integrated luminosity.} 
\label{fig:xsec}
\end{figure}

The theoretical formulas \leqn{eq:BH1} -- \leqn{eq:KKgauge} have been 
implemented in PANDORA~\cite{pandora} to
scan the scale parameters of our models (referred to as VV, AA, SR, KK+, KK$-$
and TeV in the following). At each scan point the angular distribution of the
produced leptons is studied calculating the expected number of events in
10 bins of $\cos\theta$, with a cut of $|\cos\theta|<$0.9 imposed on the
outgoing electron polar angle in the case of Bhabha scattering. The ratio of the
predicted new physics cross-section to the SM cross-section for electron and 
muon pair production is shown in
Figure~\ref{fig:xsec} 
for all the models 
considered at a 500 GeV LC.

For each considered model and parameter value
100$-$1000 Monte Carlo (MC) experiments are generated 
using Poisson statistics. These are in turn
compared to all theoretical models (including SM) by calculating the 
$\chi^2$ of the MC and the predicted theoretical distributions, 
accounting for a fully correlated systematic error of 2\% as well. 
(We will use the term {\em true model} for the model which is assumed to be
true, ie.\ which was used to generate
the MC experiments.) 
We define the confidence level (CL) at which a model with a given parameter 
can be excluded by the ratio of its $\chi^2$ probability to the highest 
$\chi^2$ probability for any model with any parameter considered:
\beq 1 - CL = {\cal P}(\chi^2) / \max({\cal P}(\chi^2))
\eeq{eq:cl}
The expected CL is computed as the median value for all the 
MC experiments generated with the same model and parameter value.

\section{Results} 

\begin{figure}[htbp]
\centerline{ 
\includegraphics[width=13cm]{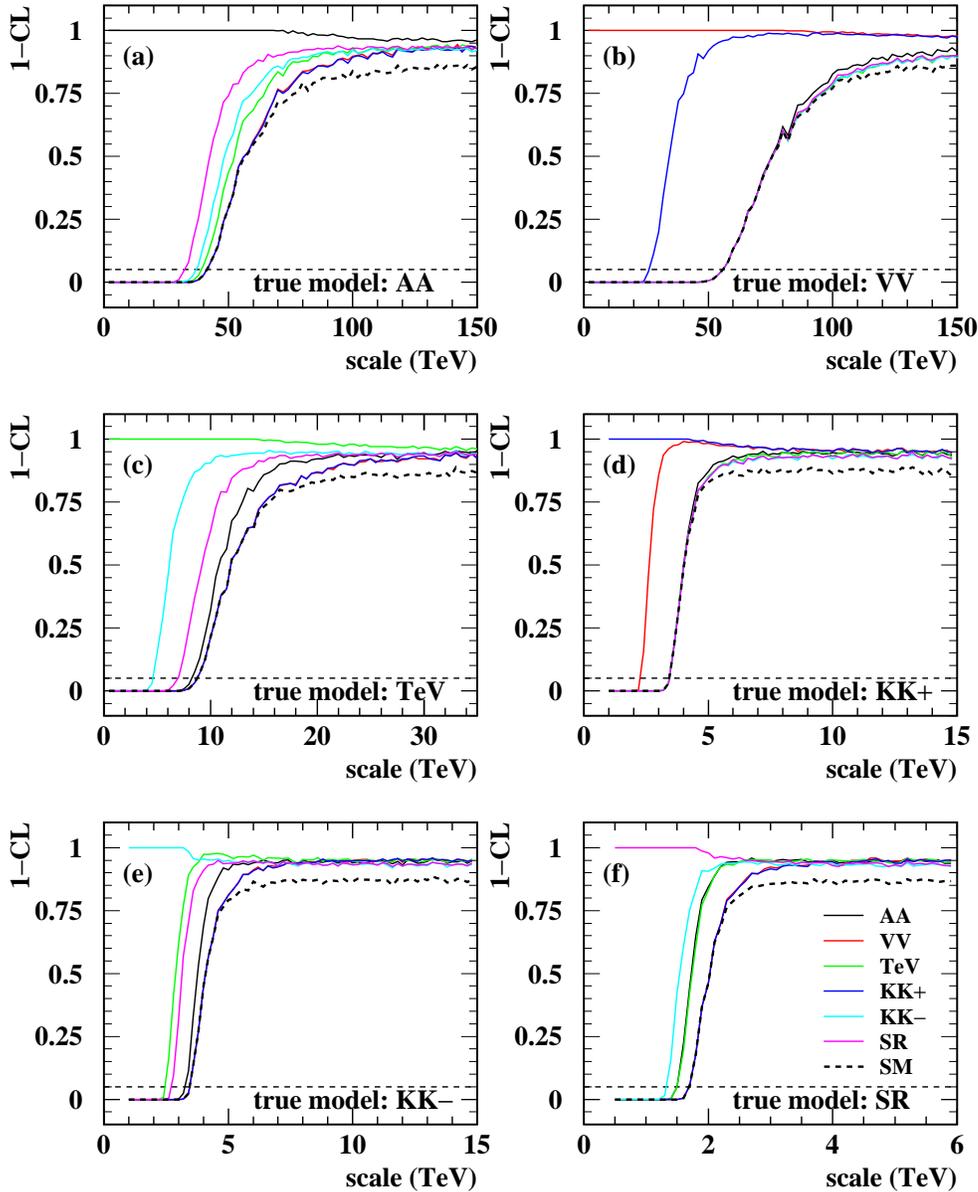}
}
\caption{Best expected $1-CL$ for each tested model,
using the electron pair final state only,
as a function of the scale parameter of the
true model  
(a) AA, 
(b) VV, 
(c) TeV, 
(d) KK+
(e) KK$-$
and
(f) SR
at a 500 GeV LC 
with an integrated luminosity of ${\cal L}=100$ fb$^{-1}$.} 
\label{fig:bestchi2}
\end{figure}

For each new physics model considered, we have calculated the 
maximum value of the scale parameter
for which the Standard Model hypothesis is expected to be excluded at the
95\% CL. We list these limits, for three sample values of the
LC energy and luminosity, in Table~\ref{tab:sm}.  
The corresponding limits for the exclusion of {\it all} models but the
true one are inevitably somewhat lower, as shown in the Table~\ref{tab:new}. 
This effect is more pronounced when only one channel is analysed, 
as is necessarily the case for the SR model where at present theoretical 
calculations only exist for Bhabha scattering. In most cases, however, 
combining all the channels allows one to distinguish between the theoretical 
models almost up to the SM sensitivity reach of Table~\ref{tab:sm}, 
with the model selection sensitivity reach of Table~\ref{tab:new} 
being only about 5-15\% lower. In Figure~\ref{fig:bestchi2} 
we plot $1-CL$ corresponding to the best fit for each tested model in the 
electron pair final state as a function of the scale parameter of the true 
model for the case of a 500 GeV LC with 100 fb$^{-1}$ luminosity.

%
\begin{table}
\caption{Highest scale parameter values of the true model 
for which the SM hypothesis 
is expected to be excluded at the 95\% CL 
for different LC energies and luminosities.
The first numbers correspond to the results using all final states 
and the second using only electron pairs.}
\label{tab:sm}
\begin{tabular}{cccc}
\hline
\multicolumn{4}{c}{scale parameter (TeV)} \\
\hline
true  & $\sqrt{s}$=500 GeV & $\sqrt{s}$=500 GeV & $\sqrt{s}$=1 TeV \\ 
model  & ${\cal L}$=100 fb$^{-1}$ 
                           & ${\cal L}$=500 fb$^{-1}$ 
			                        & ${\cal L}$=500 fb$^{-1}$ \\
\hline
AA ($\Lambda$) & 71 / 41 & 105 / 61 & 149 / 86\\
VV ($\Lambda$) & 85 / 56 & 128 / 83& 178 / 118 \\
TeV ($M_c$) & 14 / 8.5 &  21 / 13 &  29 / 18.5 \\
KK+ ($M_H$) & 3.4 / 3.4 & 4.1 / 4.1 & 7.1 / 7.0\\
KK$-$ ($M_H$) & 3.4 / 3.4 & 4.2 / 4.2  & 7.1 / 7.1\\
SR ($M_s$) & - / 1.7 & - / 2.1  & - / 3.5  \\
\hline
\end{tabular}
\end{table}

\begin{table}
\caption{Highest scale parameter values of the true model 
for which all other model hypotheses 
are expected to be excluded at the 95\% CL 
for different LC energies and luminosities.
The first numbers correspond to the results using all final states 
and the second using only electron pairs. The second best model is given in
the last column.}
\label{tab:new}
\begin{tabular}{ccccc}
\hline
\multicolumn{5}{c}{scale parameter (TeV)} \\
\hline
true  & $\sqrt{s}$=500 GeV & $\sqrt{s}$=500 GeV & $\sqrt{s}$=1 TeV & second \\ 
model  & ${\cal L}$=100 fb$^{-1}$ 
                           & ${\cal L}$=500 fb$^{-1}$ 
			                        & ${\cal L}$=500 fb$^{-1}$ 
						                   & best model \\
\hline
AA ($\Lambda$) & 68 / 32 & 101 / 48 & 142  / 70  & KK+ / SR\\
VV ($\Lambda$) & 74 / 26 & 111  / 37 & 157 / 54 & KK+\\
TeV ($M_c$) & 12 / 4.2 & 18   / 6.5 & 25.5  / 9.5 & KK$-$\\
KK+ ($M_H$) & 3.2 / 2.3 &3.9  / 2.7 & 6.7  / 4.8 & VV\\
KK$-$ ($M_H$) & 3.2  / 2.4 & 3.9 / 3.0 & 6.6  / 5.1 & TeV\\
SR ($M_s$) & - / 1.3 & - / 1.6 &- / 2.7 & KK$-$\\
\hline
\end{tabular}
\end{table}

\begin{figure}[htbp]
\vspace*{0.7cm}

\centerline{ 
\includegraphics[width=13cm]{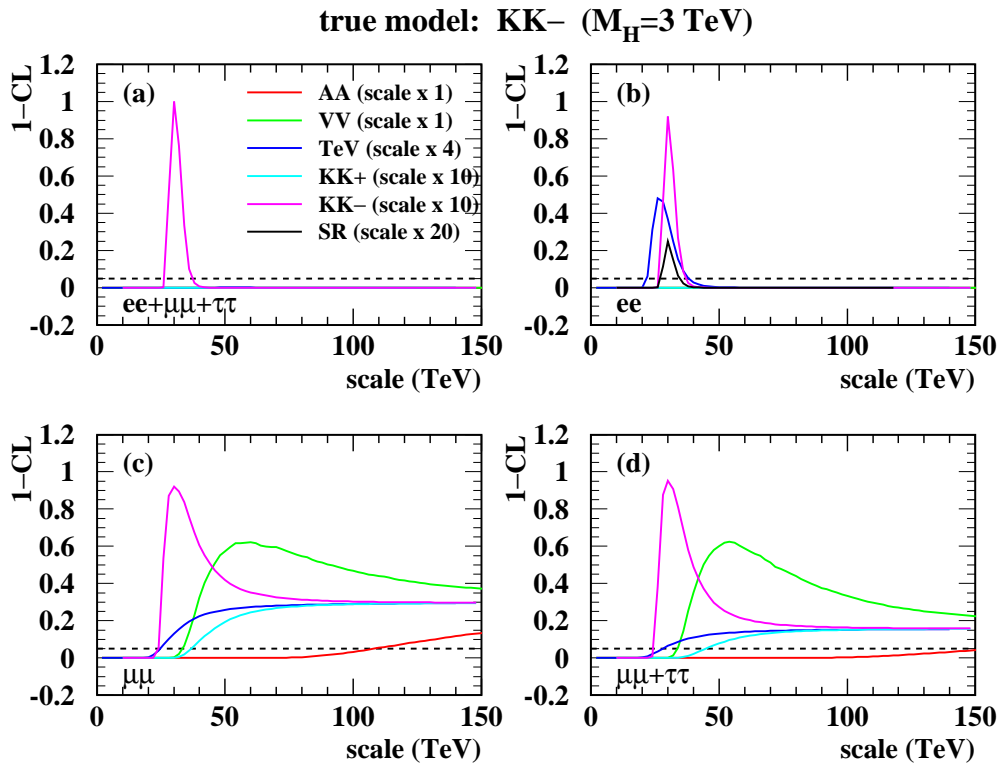}
}
\caption{Expected $1-CL$ as a function of the scale parameter of the tested model
using 
(a) all three cross-section measurements, 
(b) the electron pair, 
(c) the muon pair and 
(d) the muon and tau pair measurements 
at a 500 GeV LC 
with an integrated luminosity of ${\cal L}=100$ fb$^{-1}$. The true model 
assumed is KK$-$ with $M_H$=3 TeV.
Note that for better visibility the scale parameters have been multiplied
by 1$-$20 as indicated in the figure. 
} 
\label{fig:chi2}
\end{figure}

Measurements of electron, muon and tau final states provide complementary 
information, and combining them significantly improves model selection 
sensitivity. This is illustrated by Figure~\ref{fig:chi2}, which shows the 
$1-CL$ values as a function of the 
model scale parameter with the assumption that model KK$-$ is realized with 
$M_H$=3 TeV. While separately none of the measurements can exclude the other
models, together they do so with a high confidence level. Note that 
the sharp peak in $1-CL$ for the true model at the true parameter
value indicates that not only the model can be recognised, but the value of
its scale parameter can be estimated with a reasonable precision. Of course,
this measurement becomes less precise for higher values of the scale parameter.

\section{Conclusions}

Many models of physics beyond the Standard Model predict new particles
at the TeV scale. Even if the collider energy is not sufficient to produce
these particles directly, their virtual exchanges can still lead to observable
effects, such as contact-interaction type corrections to Standard
Model processes.   
If such a correction is observed, studying it carefully can provide 
important information about the physics at and above the TeV scale. In this 
study, 
we have considered a few well-motivated theoretical models which 
predict contact interaction corrections to lepton pair production
processes. We have shown that for a wide range of model parameters, 
measuring lepton pair production cross-sections and angular
distributions at a 500 GeV $-$ 1 TeV linear collider with realistic
integrated luminosities will allow to unambiguously determine which of the
candidate models is correct. In fact, we find that whenever a significant 
deviation from the Standard Model is seen, the model selection can be
performed with a high degree of certainty. Combining the measurements 
with electron, muon and tau final states is crucial for model selection.  


%
%

%
%


\begin{acknowledgments}
G.P. was partially supported in Snowmass by a DPF Snowmass Fellowship and by
the Hungarian Scientific Research Fund under the contract numbers OTKA T029264
and T029328. M. P. is supported by the Director, Office of Science, Office
of High Energy and Nuclear Physics, of the U. S. Department of Energy 
under Contract DE-AC03-76SF00098.
\end{acknowledgments}



\end{document}